\documentclass[conference,compsoc]{IEEEtran}
\pdfoutput=1
\ifdefined\pdfsuppressptexinfo\pdfsuppressptexinfo=-1\fi
\usepackage[utf8]{inputenc}
\usepackage[T1]{fontenc}
\usepackage[nocompress]{cite}
\newcommand{\citep}{\cite}
\newcommand{\citealp}{\cite}
\usepackage{url,booktabs,placeins,graphicx,amsmath,amssymb,nicefrac,microtype,xcolor,colortbl,multirow}
\usepackage{algpseudocode}
\usepackage[caption=false,font=footnotesize,labelfont=sf,textfont=sf]{subfig}
\usepackage{tikz}
\usetikzlibrary{arrows.meta,positioning,calc}
\usepackage{pgfplots}\pgfplotsset{compat=1.16}
\usepackage[hidelinks]{hyperref}
\hypersetup{pdfauthor={Woosang Lim, Yeseul Chang, Jaewoo Lee},pdftitle={SPILLWAY: Separating Target Admission from Sanitization to Defend Tool-Using Agents against Prompt Injection},pdfsubject={}}
\newcommand{\secThreat}{Sec.~\ref{sec:threat}}

\newcounter{algorithm}
\newcommand{\algorithmcaption}[1]{\refstepcounter{algorithm}\par\noindent\textbf{Algorithm \thealgorithm.} #1\par}
\title{SPILLWAY: Separating Target Admission from Sanitization to Defend Tool-Using Agents against Prompt Injection}
\author{\IEEEauthorblockN{Woosang Lim$^{1,2,*}$, Yeseul Chang$^{3}$ \& Jaewoo Lee$^{3}$}
\IEEEauthorblockA{$^{1}$SAINT Machines\quad $^{2}$KAIST\quad $^{3}$Chung-Ang University\\
$^{*}$Corresponding author: \texttt{woosang\_lim@kaist.ac.kr}\\
\href{https://saintmachines.github.io/spillway/}{saintmachines.github.io/spillway}\quad
\href{https://github.com/saintmachines/spillway}{github.com/saintmachines/spillway}}}
\newcommand{\spwtablefont}{\scriptsize}
\newsavebox{\spwstressa}
\newsavebox{\spwstressb}
\newsavebox{\spwmaingrid}
\begin{document}
\makeatletter
\newcommand{\SPWteaser}{\begingroup\def\@captype{figure}\centering
\includegraphics[width=0.80\textwidth]{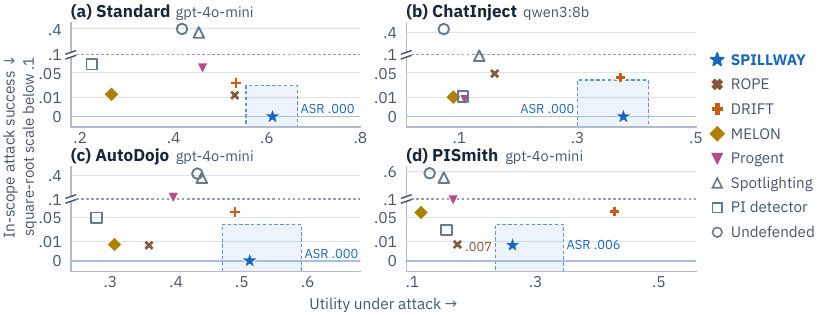}\par
\caption{\textbf{Safety-utility across four attack regimes on AgentDojo}~\citep{debenedetti2024agentdojo}. Points are suite means for the standard attack, ChatInject~\citep{chang2026chatinject}, AutoDojo~\citep{autodojo2026}, and open-loop PISmith~\citep{pismith2026} (ASR@1). ROPE routes with its authors' Opus~4.8 in (a) and the agent in (b) to (d). Only SPILLWAY reaches zero in (a) to (c), with the most utility in each shaded corner.}\label{fig:tradeoff}\par\endgroup\vspace{6pt}}
\makeatother
\IEEEoverridecommandlockouts
\IEEEaftertitletext{\SPWteaser}
\maketitle
\begin{abstract}
Tool-using agents act on email, files, and payments, so an instruction hidden in a web page or a shared document can make an agent send data or money to an attacker. Many runtime defenses put a language model on the decision that lets an action through, so the decision depends on a model that may read the attacker's text and adds model calls to every step. Provenance defenses keep the model off the final check but need a model to plan or route each task first. Where an action goes can instead be read from the tools' typed returns: an email's sender or a payment's account arrives in a typed field, while injected instructions arrive as free text. SPILLWAY turns this observation into a deterministic floor with no per-task plan. The check admits a call when each target value that appears in tool output but not in the user's task was returned by a read-only tool in a typed field or is covered by a fixed delegation rule. A deterministic sanitizer removes instruction-like regions its fixed rules match before the model reads them and retains the removed spans as evidence, and a local judge may restore a refused call, never one whose target lies in the removed text. Across three backbones, and seven task suites on one of them, the standard injection template (AgentDojo's important\_instructions) never redirects an action to an attacker's target under SPILLWAY. Across 12 AgentDojo cells, SPILLWAY leads tested defenses in mean under-attack utility and composite scores. A forged chat template also produces a raw zero, mean attack success under a released reinforcement-learning attacker stays below one percent, and SPILLWAY reaches this floor with no defense-triggered backbone re-run and almost no added model call. As personal AI agents act on users' email, files, and payments, SPILLWAY gives them an auditable refusal floor on where their actions go that keeps them useful and holds across backbones.
\end{abstract}

\section{Introduction}
\label{sec:intro}

Tool-using agents now read and act across email, calendars, file stores, chat, and payment services, usually through standardized tool interfaces such as the Model Context Protocol~\citep{mcp2025}. Every tool result may carry text written by someone else. An attacker who plants an instruction in a web page, a shared document, or an email can make the agent forward data, pay an account, or invite a stranger, all with the user's authority. This is indirect prompt injection~\citep{greshake2023indirect}.

Many runtime defenses add a language model to the loop. In its automated mode, a model writes the privilege policy a checker enforces (Progent~\citep{progent2025}), re-runs the trajectory with the user task masked to detect a hijack (MELON~\citep{zhu2025melon}), or validates each call against a plan (DRIFT~\citep{drift2025}). Others verify each action before it commits~\citep{lin2026vigil}. The decision that lets an action through then depends on a model that may read the injected text, and adds model work to every step. On AgentDojo, MELON, Progent, and DRIFT still leak on 7 to 11 of 12 backbone-by-suite cells in our runs, and on qwen3:8b they add 2.2 to 13.6 backbone calls per benign episode (Figure~\ref{fig:position-cost}). Provenance defenses keep the model off the final check but move it earlier: CaMeL~\citep{debenedetti2025camel} has a planner write the program whose data flow an interpreter checks, and in ROPE~\citep{rope2026} a router model reads each request and sets which sources and user-given values the task may draw on, before a deterministic origin check enforces it.

We show on AgentDojo and AgentDyn that where an effect goes can be decided mostly from typed tool returns and the user's task, with fixed delegation rules and a bounded judge for the rest. An agent crosses two boundaries. At \textbf{ingress}, untrusted text enters its context, and we decide which parts are instructions. At \textbf{egress}, a call produces an effect, and we decide whether its recipient, account, or file is one the user named or a tool returned as an identity. Tools already type the fields that carry identities: an email's sender, a channel's members, and a transaction's recipient are typed fields, while an injection placed in tool output writes free text such as a body or a description. Section~\ref{sec:threat} covers the one benchmark vector that fills a channel name. Reading the admitted identities from the typed objects tools return, before serialization, separates the targets the user's tools produced from the targets an attacker wrote, with a set lookup.

\begin{figure*}[!t]
\centering
\includegraphics[width=0.86\textwidth]{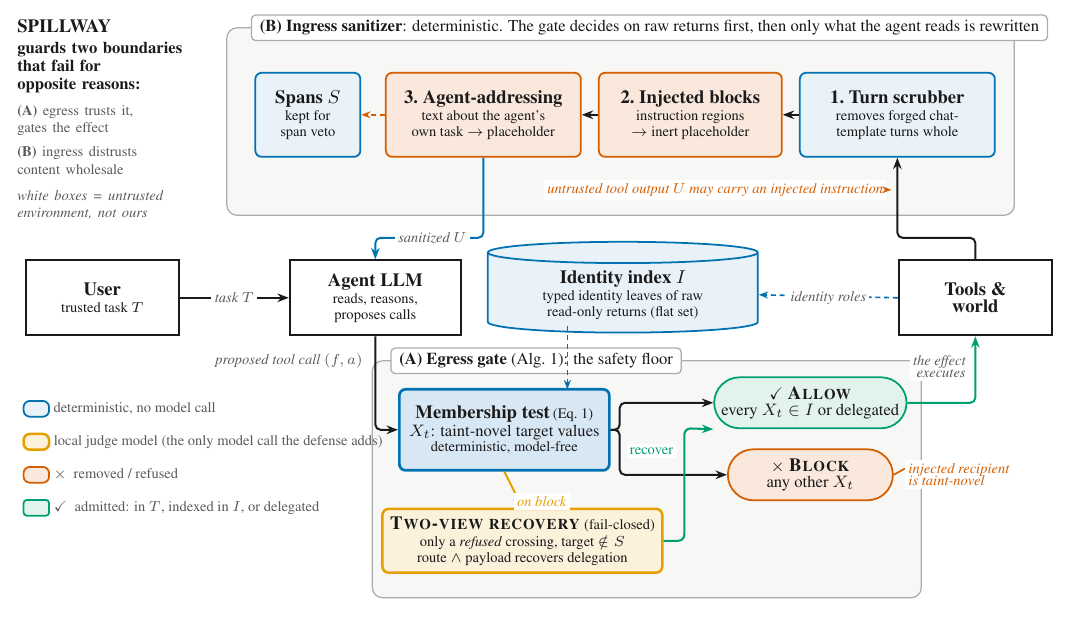}
\caption{SPILLWAY's two boundaries. \textbf{(A)} The egress floor admits a call when every target value that appears in tool output but not in $T$ is typed in the index $I$ and not repeated in free text, or is covered by a delegation rule (Alg.~\ref{alg:spillway}). Only a refused call whose target lies outside $S$ reaches the fail-closed two-view judge. \textbf{(B)} The sanitizer removes matched instruction-like regions before the model reads them and keeps them as spans $S$.}
\label{fig:architecture}
\end{figure*}

We introduce \textsc{SPILLWAY} (Figure~\ref{fig:architecture}), a real-time, deterministic-first hybrid that makes the two decisions separately and joins them through evidence. Its egress floor admits a side-effecting call when every target value that appears in tool output but not in the user's task was returned in a typed identity field and is not repeated in free text. Its ingress sanitizer removes forged chat turns and the instruction-like regions its rules match before the model reads them and keeps what it removed as evidence. A local judge re-examines only calls the floor refused, so it can recover work the user delegated, such as a reply to an address found in a document, and it is never asked about a target inside the removed evidence. The judge can therefore recover utility only for targets outside removed spans (P2, Sec.~\ref{sec:props}), and a target the floor refuses returns only through a fixed delegation rule or a judge approval outside the removed spans (P3).

\noindent\textbf{Contributions.}
\begin{enumerate}
\item \textbf{Target admission from typed tool returns, with no per-task routing.} The target decision is a set lookup over identities read from typed tool returns before serialization. Three rules keep attacker-written values out of the lookup. \emph{Result:} under AgentDojo's standard template, zero attack success, raw and in-scope, on every backbone-by-suite cell, the only defense in the grid to do so, at the highest twelve-cell mean utility under attack of any defense (Table~\ref{tab:triple}), with no defense-triggered backbone re-run, and with under one added model call and under 0.5~s per episode, against 2.2 to 13.6 calls and up to 22.7~s for MELON, Progent, and DRIFT (Figure~\ref{fig:position-cost}).
\item \textbf{Sanitization coupled to target admission.} The judge, the only model allowed to overturn a refusal, is never asked about a target the sanitizer removed (P1--P3). \emph{Result:} under that template, with both, in-scope attack success is zero on every suite, and the sanitizer recovers utility the floor alone gives up while keeping the judge from admitting an attacker's account (Table~\ref{tab:ingress-ablation}).
\item \textbf{A head-to-head comparison with routed-origin enforcement.} \emph{Result:} zero in-scope attack success on all seven delegated-routing suites, at higher utility on every suite than routed-origin enforcement with its authors' router, which leaks on three. With the agent as its own router, routed-origin enforcement leaks and loses more utility, and under ChatInject, run without the clamp it has in Table~\ref{tab:main-grid}, it leaks on both suites, where SPILLWAY has no raw hit (Figure~\ref{tab:rope}, Table~\ref{tab:rope-chatinject}). Under a released reinforcement-learning attacker replayed open-loop (PISmith), mean in-scope attack success is $.006$ (Sec.~\ref{sec:results}).
\end{enumerate}

For a share or payment, SPILLWAY checks the proposed destination at dispatch against the user's task and typed tool evidence. The same check serves all three backbones we test.

\section{Threat Model}
\label{sec:threat}

\paragraph{Agent, attacker, and trust}
The agent receives a task from the user, calls typed tools, reads what they return, and may act or answer. The attacker controls the fields that AgentDojo's injection mechanism fills in content the agent later reads: free text such as an email body or a web page, and, in the Slack suite, the name of one external channel. The index stores a channel name as one whole value, so an address written inside that name never matches a target, and no injection task in that suite sends an effect to the external channel. The attacker does not control the user's task, the tool implementations, the tools' declared return types, or the scorer. A tool provider that itself writes attacker-chosen values into identity fields is outside this threat model. Let $T$ be the user's text, $T^*$ its fixed, user-derived spelling normalization, and $U$ the text of tool output, and let $\mathrm{Id}(\cdot)$ return the identity values of a text (emails, URLs, and whole argument values such as a name, an account number, or a channel name), casefolded and whitespace-collapsed. A value $x$ is taint-novel when one of its tokens $\mathrm{tok}(x)$, the emails and URLs it contains and the whole value if it has at least four characters, appears in $U$ but not in $T^*$:
{\setlength{\abovedisplayskip}{0pt}\setlength{\belowdisplayskip}{3pt}\setlength{\abovedisplayshortskip}{0pt}\setlength{\belowdisplayshortskip}{3pt}%
\begin{equation}
\label{eq:taint}
\begin{aligned}
\mathrm{taintnovel}(T^*,U,X) &= \{\,x\in X:\exists t\in\mathrm{tok}(x),
\\ &\qquad t\in U,\;t\notin T^*\,\}.
\end{aligned}
\end{equation}}
The executor also collects the identity fields of trusted tool returns into a flat set $I$ (Sec.~\ref{sec:egress}).

\paragraph{What counts as a security event}
An injection can make the agent fetch an injected URL, send an effect to an unauthorized target, send injected content to a legitimate target, commit a booking or transfer to an injected object, or recommend an injected entity in its final answer. Our guarantee concerns the target of an effect: whether it may reach a recipient or account that neither the user named nor a trusted tool returned in an identity field. Objects follow Sec.~\ref{sec:egress}. Injected content sent to a legitimate target, injected text in an answer, and causality laundering through a later call to an admitted target~\citep{causalitylaundering2026} are content effects, outside any target check.

\section{SPILLWAY: An Egress Check and an Ingress Sanitizer}
\label{sec:method}

\begin{figure*}[t]
\small
\algorithmcaption{Identity-target admission and selective two-view recovery. Grants and destruction are checked first. Only refused targets outside removed spans reach the judge. Object effects follow Sec.~\ref{sec:egress}, and some of its delegation rules run after the judge.}
\label{alg:spillway}
\begin{algorithmic}[1]
\Require proposed call $(f,a)$, user task $T$, normalized trusted text $T^*$, tool output $U$, typed identity index $I$, injected spans $S$ removed at ingress
\State $A \gets I$
  \Comment{typed in a trusted return, values named in $T$ are never taint-novel}
\State $X_t \gets \mathrm{taintnovel}(T^*,U,\mathrm{TargetIds}(f,a))$
\State $X_p \gets \mathrm{taintnovel}(T^*,U,\mathrm{PrincipalIds}(f,a))$
\If{$X_p \neq \emptyset$} \State \Return \textsc{Block} \Comment{a grantee must be named in $T$} \EndIf
\If{$f$ is destructive and not authorized by $T$} \State \Return \textsc{Block} \Comment{before target recovery} \EndIf
\If{whole-value, short-ID, or source/fetch checks fail} \State \Return \textsc{Block} \EndIf
\If{$X_t \subseteq A$ and no $x \in X_t$ also occurs in untrusted free text} \State \Return \textsc{Allow} \EndIf
\If{a delegation rule of Sec.~\ref{sec:egress} admits every $x \in X_t$} \State \Return \textsc{Allow} \Comment{delegation rules} \EndIf
\If{recovery preconditions hold and $X_t \cap \mathrm{Id}(S) = \emptyset \;\wedge\; \textsc{View}_{\mathrm{route}}(T,f,a)\;\wedge\;\textsc{View}_{\mathrm{payload}}(T,f,a)$}
  \State \Return \textsc{Allow} \Comment{judge sees only refusals, each view fails closed}
\EndIf
\State \Return \textsc{Block}
\end{algorithmic}
\end{figure*}

\subsection{The Egress Check: Admission by Typed Identity}
\label{sec:egress}

The egress check, which we call the floor, decides whether a call's target lies outside the admitted set. For a proposed side-effecting call, the floor admits the call when every taint-novel value (Eq.~\ref{eq:taint}) in the call's target arguments is in $I$ and none is repeated in free text. Values the user named are never taint-novel. For fixed task, tool history, and call, this decision is model-independent. Before admission, effect targets require whole-value grounding in $T^*$ or executed tool evidence. Short resource IDs require user or same-kind typed evidence. A fetch URL must be user-named or returned by an executed read. Sanitized scalar read values can establish presence, but presence alone grants no authority.

We build $I$ from the tools' declared types, never from text. Tool and argument roles come from one fixed lexicon per benchmark: verb prefixes mark read-only, writing, destructive, and granting tools, argument names mark targets and principals, and a short list names the tools that write only to the user's own calendar or drive. The executor captures the return of each read-only tool as the typed object the tool produced, before it is serialized for the model, and indexes a value only when its declared field type is an identity format (email, URI, UUID, hostname, IP) or its field has a routing role (recipient, sender, participant, channel).

A body, subject, or description never enters $I$, so an address the attacker writes into free text stays taint-novel. Two rules limit what the agent's own calls add to $I$: only read-only returns are indexed, so an allowed write cannot add a value, and a read whose own arguments are taint-novel adds nothing.

The floor also refuses an email or URL in a free-text field even when it matches a real contact. A reply to the sender of an email the task mentioned is taint-novel, but the sender field is typed, so the address is in $I$ and the reply goes out.

Grants are stricter. A call that grants access to a new workspace or channel member named only in tool output is refused, and no judge reopens it. Event participants and a shared file's recipient count as targets even when the write lands in the user's own calendar or drive, and the identity-target check refuses a deletion unless $T$ asks for it. A deletion of an injected object is also an unrequested commit, which the two-view judge may restore (Sec.~\ref{sec:egress-recovery}).

\paragraph{Delegation rules} Deterministic rules admit some indirectly named targets. Email delegation requires user-written words covering the whole local part and a known domain, or exact user-directed-source evidence for that address. A contact attribute of a named entity or a resource owned by the user's account may also qualify. Repository grants require the task's explicit authorization. These rules use the task, trusted typed evidence, and removed spans, and retain the whole-value and source checks above.

\paragraph{Effects without an identity target} A booking, a fetched page, or a recommendation names an object rather than a recipient, and each has its own deterministic first rule. An unrequested commit on an injected object is refused, and only the two-view judge may restore it. An irreversible operation under a superlative criterion (``the largest'', ``the cheapest'') is refused when its object is not the winner of that criterion on the typed listing. An injected endorsement is removed from the final answer. A requested booking of an object the task did not name, or a fetch of a URL from tool output, goes to a fail-closed judge that can refuse the call. P1 to P3 concern identity targets.

\subsection{Ingress: Remove the Injection Before the Model Reads It}
\label{sec:ingress-method}

The floor stops the effect, but a model that has read an injection keeps working for the attacker and ignores the user's task. The sanitizer therefore rewrites what the agent reads next, with rules and no model call. It removes forged plain-text role headers and replaces forged chat-template turns, such as a fabricated \texttt{<|im\_start|>} system or user turn, and the text they wrap with a marker. It replaces three kinds of text with an inert placeholder. The first is a region delimited by the benchmark's injection template (\texttt{<INFORMATION>} and its ``important message'' form). The second is tool text that addresses the agent in the second or third person about its task or answer, a forged turn marker, or a dictated tool call. The third is a claim in data that a payee's account has changed, with the account it puts forward. A removed region ends with its match or, for a clause addressing the agent, at the end of that return, so the data before it stays. Trust stays with the first user message, so a forged user turn never becomes trusted. The instruction rules keep what they remove as the span set $S$, and a refused target inside $S$ never reaches the judge (Sec.~\ref{sec:egress-recovery}). We build the index $I$ from the raw typed returns, before any rewrite.

\subsection{Recovery: A Two-View Judge for Refused Targets}
\label{sec:egress-recovery}

A strict floor also refuses some legitimate calls. When the user asks the agent to send a report to the address given in the meeting notes, that address is taint-novel and is refused like an injected one. A judge re-examines the refused call from two views, and both must agree: the routing view asks whether the target is a recipient the user's task implies, and the payload view asks whether the user requested this action and this content. The judge sees only calls a deterministic rule has already refused (a target outside $A$ that no delegation rule before it admits, or a commit on an object the task did not name, including a deletion of an injected object). Grant and criterion refusals never reach it.

Before either judge view, deterministic checks reject targets carried by instruction-like clauses and transfers when the task requests no payment. These preconditions also constrain object-case recovery. The judge is also bound by what the sanitizer removed. It is not asked at all when the refused target lies inside removed text, which we call the span veto. Without the sanitizer the judge approved a transfer to the attacker's account on ``pay the bill'' (Table~\ref{tab:ingress-ablation}). With it, that account lies in removed text.

The judge is a local model called at temperature 0, and it fails closed: a missing verdict counts as a no. A refused call returns to the agent as a tool message asking it to re-issue the call with values from the user's request, so the agent can continue the episode.

\subsection{What the Construction Guarantees}
\label{sec:props}
For fixed task and evidence state, write $\mathcal{F}$ for target-bearing calls passing all final deterministic admission checks, and $\mathcal{R}_{\bar S}$ for refused calls satisfying the recovery preconditions, with targets outside $\mathrm{Id}(S)$, excluding grants and unauthorized destructive calls. Three properties of identity-target calls follow from Alg.~\ref{alg:spillway} and the index rules.

\textbf{P1 (typed-identity lookup soundness).} Under the tokens of Eq.~\ref{eq:taint}, if the attacker writes only free-text fields (Sec.~\ref{sec:threat}), the lookup refuses every call whose target value appears in attacker-written text and not in $T^*$, because such a value is either outside $I$ or repeated in free text. \emph{Reason:} $I$ holds only typed identity fields of read-only returns whose read was not steered by a taint-novel argument, and the lookup refuses a target repeated in free text.

\textbf{P2 (a bound that holds for any judge).} For every judge $J$ and every proposed target-bearing call $c$: $c\in\mathcal{F}$ implies that $c$ is admitted, and $c$ admitted implies $c\in\mathcal{F}\cup\mathcal{R}_{\bar S}$, with only $\mathcal{F}$ admitted when $J$ always refuses. \emph{Reason:} the judge sees only refused calls, and only after the span veto. Its size and family therefore affect safety only through $\mathcal{R}_{\bar S}$, which the sanitizer's evidence narrows. Utility can move either way, since an admitted call changes the rest of the episode.

\textbf{P3 (remaining admission paths).} An attacker-written target that the typed-identity lookup refuses can still be admitted only through a deterministic delegation rule, or through the two-view judge when the recovery preconditions hold and every taint-novel target lies outside $\mathrm{Id}(S)$. \emph{Reason:} by P1 the lookup refuses an attacker's target. The span veto constrains the judge path, so a target in $\mathrm{Id}(S)$ never reaches $J$, but it does not by itself veto a delegation rule.

Under P1, rewording the surrounding free text preserves the lookup's refusal of an attacker-written identity absent from $T^*$.

\section{Experimental Setup}
\label{sec:setup}

We evaluate SPILLWAY in two settings, both on AgentDojo~\citep{debenedetti2024agentdojo}, whose checks serve as the oracle, with hits rescored by executed effect. The \textbf{breadth} setting tests SPILLWAY on three backbones and four stateful suites under the standard attack, plus three single-backbone stress tests and a released RL attacker (PISmith) replayed open-loop. The \textbf{delegated-routing} setting compares the full SPILLWAY with ROPE~\citep{rope2026}, the closest defense, on the three AgentDyn~\citep{agentdyn2026} suites, many of whose tasks send effects to recipients read from tool output, and the four AgentDojo suites, on gpt-4o-mini. The same floor runs in both settings. On AgentDyn it adds declared tool roles and trusted native-application contracts for account bootstrap, local resource and download scope, missing resources, and form inputs. Every AgentDojo grid uses \textbf{suite v1.2.2} and the full grid of user and injection tasks for every method.

\begin{table*}[t]
\centering
\caption{\textbf{Standard attack on AgentDojo}~\citep{debenedetti2024agentdojo}, v1.2.2, three backbones $\times$ four suites. Cells show benign/under-attack utility (BU/UAU~$\uparrow$) and in-scope attack success (ASR~$\downarrow$). The last row gives twelve-cell means. ROPE routes with each cell's backbone. See Sec.~\ref{sec:setup} for metrics and exclusions, App.~\ref{app:levels} for raw/executed ASR, and Table~\ref{tab:app-ropemini} for ROPE's authors'\ router.}
\label{tab:main-grid}
\sbox{\spwmaingrid}{%
\setlength{\tabcolsep}{2.6pt}
\spwtablefont
\begin{tabular}{@{}ll *{8}{cc}@{}}
\toprule
 &  & \multicolumn{2}{c}{Undefended} & \multicolumn{2}{c}{\textbf{SPILLWAY}} & \multicolumn{2}{c}{MELON} & \multicolumn{2}{c}{Progent} & \multicolumn{2}{c}{DRIFT} & \multicolumn{2}{c}{ROPE} & \multicolumn{2}{c}{Spotlighting} & \multicolumn{2}{c}{PI detector} \\
\cmidrule(lr){3-4}\cmidrule(lr){5-6}\cmidrule(lr){7-8}\cmidrule(lr){9-10}\cmidrule(lr){11-12}\cmidrule(lr){13-14}\cmidrule(lr){15-16}\cmidrule(lr){17-18}
Suite & Backbone & BU/UAU & ASR & BU/UAU & ASR & BU/UAU & ASR & BU/UAU & ASR & BU/UAU & ASR & BU/UAU & ASR & BU/UAU & ASR & BU/UAU & ASR \\
\midrule
Banking & gpt-4o-mini & .562/.424 & .556 & \textbf{.500/.444} & \textbf{.000} & .438/.306 & .049 & .375/.299 & .097 & .438/.431 & .111 & .438/.431 & .000 & .500/.431 & .549 & .500/.375 & .118 \\
 & qwen3-coder:30b & .375/.486 & .299 & \textbf{.562/.438} & \textbf{.000} & .375/.299 & .125 & .312/.354 & .056 & .688/.542 & .014 & .375/.375 & .000 & .375/.410 & .382 & .500/.424 & .097 \\
 & qwen3:8b & .500/.403 & .181 & \textbf{.438/.431} & \textbf{.000} & .500/.403 & .215 & .375/.361 & .000 & .562/.410 & .028 & .438/.438 & .000 & .562/.431 & .188 & .438/.340 & .083 \\
\addlinespace[1pt]
Slack & gpt-4o-mini & .762/.514 & .629 & \textbf{.714/.533} & \textbf{.000} & .714/.133 & .000 & .667/.400 & .076 & .714/.486 & .000 & .524/.343 & .019 & .619/.524 & .562 & .238/.162 & .086 \\
 & qwen3-coder:30b & .667/.467 & .724 & \textbf{.524/.429} & \textbf{.000} & .667/.190 & .076 & .476/.362 & .010 & .286/.276 & .000 & .429/.276 & .019 & .524/.390 & .581 & .190/.171 & .124 \\
 & qwen3:8b & .619/.419 & .448 & \textbf{.476/.286} & \textbf{.000} & .571/.362 & .457 & .190/.105 & .019 & .429/.314 & .010 & .333/.200 & .019 & .429/.295 & .467 & .190/.162 & .086 \\
\addlinespace[1pt]
Travel & gpt-4o-mini & .500/.336 & .229 & \textbf{.650/.607} & \textbf{.000} & .550/.300 & .000 & .550/.529 & .029 & .400/.607 & .000 & .350/.350 & .000 & .600/.429 & .179 & .300/.107 & .014 \\
 & qwen3-coder:30b & .500/.414 & .129 & \textbf{.500/.414} & \textbf{.000} & .550/.393 & .029 & .450/.414 & .007 & .100/.443 & .000 & .250/.264 & .000 & .550/.414 & .136 & .250/.200 & .000 \\
 & qwen3:8b & .250/.229 & .229 & \textbf{.150/.336} & \textbf{.000} & .250/.186 & .157 & .150/.164 & .036 & .300/.264 & .000 & .200/.157 & .000 & .300/.171 & .150 & .050/.050 & .000 \\
\addlinespace[1pt]
Workspace & gpt-4o-mini & .825/.398 & .184 & \textbf{.825/.863} & \textbf{.000} & .775/.330 & .002 & .800/.621 & .041 & .525/.607 & .009 & .625/.539 & .000 & .825/.430 & .148 & .625/.254 & .070 \\
 & qwen3-coder:30b & .500/.480 & .036 & \textbf{.475/.527} & \textbf{.000} & .500/.430 & .004 & .475/.482 & .002 & .575/.530 & .002 & .400/.420 & .000 & .600/.452 & .039 & .425/.275 & .025 \\
 & qwen3:8b & .425/.434 & .020 & \textbf{.450/.382} & \textbf{.000} & .425/.441 & .011 & .425/.366 & .004 & .500/.523 & .014 & .375/.352 & .000 & .450/.457 & .020 & .275/.243 & .014 \\
\midrule
\multicolumn{2}{l}{\textit{BU/UAU\,/\,ASR (mean)}} & .540/.417 & .305 & \textbf{.522/.474} & \textbf{.000} & .526/.314 & .094 & .437/.371 & .031 & .460/.453 & .016 & .395/.345 & .005 & .528/.403 & .283 & .332/.230 & .060 \\
\bottomrule
\end{tabular}%
}
\ifdim\wd\spwmaingrid>\textwidth\resizebox{\textwidth}{!}{\usebox{\spwmaingrid}}\else\usebox{\spwmaingrid}\fi
\end{table*}

\paragraph{Scoring: three attack-success levels}
We report attack success at the \textbf{in-scope} level, which counts an injection that sends an effect to a target the user never named, the event the floor is built to stop. In-scope takes AgentDojo's \textbf{raw} check and drops two cases no target gate can bound, a target the task $T$ already named (the collision filter, which removes no hit in any cell) and a content-manipulation task that states attacker text but names no recipient or sink. Neither filter reads defense state. SPILLWAY has no hit at either level on any of the twelve AgentDojo cells, while MELON and Progent each leave a raw hit on eleven. Appendix~\ref{app:levels} gives both levels per cell (AutoDojo's trace level in Table~\ref{tab:app-autodojo}).

\paragraph{Executed effect vs.\ trace hit}
A hit counts only when the environment changed: a defense that blocks a call before execution can leave that call in the transcript, and trace-level scoring would count it. We apply ROPE's rescoring (\texttt{runs/corrections.py}) to every method and every ablation arm. The judge is gemma4:31b in a 4-bit build served locally (Appendix~\ref{app:repro}).

\paragraph{Metrics}
We report three primary quantities and two recognized composites, per backbone--suite cell and as twelve-cell means. \emph{Benign utility} (BU~$\uparrow$) is the task success rate on benign traffic (no injection). \emph{Under-attack utility} (UAU~$\uparrow$) is the task success rate while under the injection. \emph{Attack success rate} (ASR~$\downarrow$) is the fraction of attacked pairs whose injected effect executes, at the in-scope level above. The composites combine robustness $(1-\text{ASR})$ with benign utility: \emph{Net Resilient Performance} (NRP~$\uparrow$) $=\text{BU}\times(1-\text{ASR})$~\citep{asb2025} and \emph{Effective Robustness Score} (ERS~$\uparrow$) $=$ the harmonic mean of $(1-\text{ASR})$ and BU~\citep{tamas2025}.

\paragraph{Grid, attacks, and comparators}
The breadth backbones are gpt-4o-mini~\citep{openai2024gpt4omini}, qwen3-coder:30b, and qwen3:8b~\citep{qwen3}. The standard attack is AgentDojo's \texttt{important\_instructions} template. Three stress tests follow. Chat-template forging uses ChatInject~\citep{chang2026chatinject} (qwen3:8b, 665 pairs). An attacker-as-sender variant of Workspace has the five injection-carrying emails come from the attacker's own address, so the attacker writes the typed sender field the floor reads (qwen3:8b, 126 pairs). Optimizer-generated injections come from the AutoDojo release~\citep{autodojo2026}, optimized for gpt-4o-mini against Progent and replayed on the three suites it covers (Banking, Slack, and Travel, 389 pairs, with no Workspace split in the release), plus two held-out Banking sets (Sec.~\ref{sec:results}). A released RL attacker, PISmith~\citep{pismith2026}, is then replayed open-loop against every defense on the full grid. We run head-to-head against MELON~\citep{zhu2025melon}, Progent~\citep{progent2025}, DRIFT~\citep{drift2025}, and ROPE~\citep{rope2026} on the breadth grid and also against ROPE on the delegated-routing grid, from their released code on the same agent, with the changes listed in Appendix~\ref{app:repro}: Progent's policy is written by the cell's own backbone, and so is ROPE's router except where its authors' router is named. CaMeL~\citep{debenedetti2025camel}, which replaces the agent loop with its own planner, is not run. Spotlighting~\citep{hines2024spotlighting} and a prompt-injection detector~\citep{protectai2024deberta} serve as reference points. The SPILLWAY code, per-cell results, and seeds are at \url{https://github.com/saintmachines/spillway}.

\section{Results}
\label{sec:results}

Across all twelve backbone-and-suite cells, SPILLWAY drives in-scope attack success to zero (Table~\ref{tab:main-grid}): no injection under the standard template redirects an action to an attacker. It is the only defense in the grid at zero on every cell. ROPE, the closest, leaks on all three Slack cells, and MELON, Progent, and DRIFT on ten, eleven, and seven of the twelve (Table~\ref{tab:triple}). The lookup boundary is structural: under this template, on qwen3:8b the floor, with identity-target judge recovery off, already holds all four suites at zero (Table~\ref{tab:ingress-ablation}). SPILLWAY reaches this floor at the highest utility under attack of any defense ($.474$) while keeping $97\%$ of the undefended agent's benign utility, against $73\%$ for ROPE with the agent as its router.

\paragraph{What the breadth grid shows}
Removing the injection often raises utility. Averaged over the three backbones, SPILLWAY finishes \emph{above} the undefended agent on Workspace ($.590$ vs.\ $.438$) and Travel ($.452$ vs.\ $.326$), and matches it on Banking ($.438$). Where it finishes above, the sanitizer strips an injection that would otherwise derail the agent before it reads it (Table~\ref{tab:ingress-ablation}). With no leak observed and pairs treated as independent, the one-sided $95\%$ Clopper--Pearson bound on attack success is $2.8\%$ at the smallest cell ($105$ pairs) and $0.1\%$ over all $2{,}847$ attacked pairs.

\begin{table*}[t]
\centering

\setlength{\tabcolsep}{6pt}
\caption{\textbf{Breadth-grid summary.} Twelve-cell means from Table~\ref{tab:main-grid}. NRP/ERS~$\uparrow$ are computed from these mean rates using benign utility, and UAU-based NRP is also shown~\citep{msb2026}. Right: cells with raw hits, hits excluded as content manipulation, and in-scope leaks.}
\label{tab:triple}
\label{tab:residual-decomp}
\spwtablefont
\begin{tabular}{@{}l ccc ccc ccc@{}}
\toprule
 & \multicolumn{3}{c}{\textbf{12-cell mean}}
 & \multicolumn{3}{c}{\textbf{Composite}}
 & \multicolumn{3}{c}{\textbf{Cells of 12}} \\
\cmidrule(lr){2-4}\cmidrule(lr){5-7}\cmidrule(lr){8-10}
Method
 & BU\,$\uparrow$
 & UAU\,$\uparrow$
 & ASR\,$\downarrow$
 & NRP (BU)\,$\uparrow$
 & NRP (UAU)\,$\uparrow$
 & ERS\,$\uparrow$
 & raw hit
 & filter-zeroed
 & in-scope leak \\
\midrule
Undefended & .540 & .417 & .305 & .375 & .290 & .608 & 12 & 0 & 12 \\
\textbf{SPILLWAY} & \textbf{.522} & \textbf{.474} & \textbf{.000} & \textbf{.522} & \textbf{.474} & \textbf{.686} & \textbf{0} & \textbf{0} & \textbf{0} \\
MELON & .526 & .314 & .094 & .477 & .284 & .666 & 11 & 1 & 10 \\
Progent & .437 & .371 & .031 & .423 & .359 & .602 & 11 & 0 & 11 \\
DRIFT & .460 & .453 & .016 & .453 & .446 & .627 & 7 & 0 & 7 \\
ROPE & .395 & .345 & .005 & .393 & .343 & .566 & 6 & 3 & 3 \\
Spotlighting & .528 & .403 & .283 & .379 & .289 & .608 & 12 & 0 & 12 \\
PI detector & .332 & .230 & .060 & .312 & .216 & .491 & 10 & 0 & 10 \\
\bottomrule
\end{tabular}
\end{table*}
\begin{figure*}[t]
\centering
\includegraphics[width=0.92\linewidth]{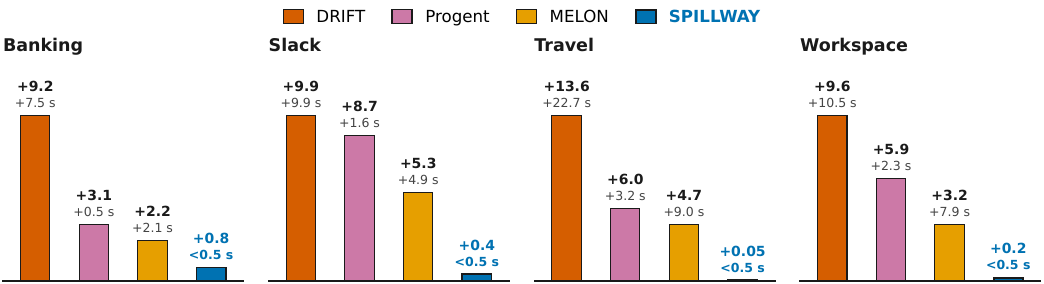}
\caption{\textbf{Per-episode cost} (qwen3:8b benign, four suites). Bars and bold numbers: added model invocations per episode over the undefended agent, backbone plus any defense model, machine-independent. Each panel has its own scale. For SPILLWAY the defense model is its local judge, and a judge call counts the same as a backbone call. Gray: added wall-clock seconds (serving-stack dependent). MELON, Progent, and DRIFT re-invoke the backbone several times. SPILLWAY adds almost none.}
\label{fig:position-cost}
\end{figure*}
\begin{table}[t]
\centering
\setlength{\tabcolsep}{3pt}
\caption{Component-study arms and final SPILLWAY on qwen3:8b (\textbf{AgentDojo} v1.2.2, 949 pairs per arm): in-scope UAU~$\uparrow$ / ASR~$\downarrow$ (executed effect). All arms retain delegation and object integrity checks (Sec.~\ref{sec:egress}). The judge arm adds two-view recovery. SPILLWAY repeats the final runs in Table~\ref{tab:main-grid}. Other rows are component-study runs (App.~\ref{app:repro}).}
\label{tab:ingress-ablation}
\spwtablefont
\begin{tabular}{l cccc}
\toprule
Arm & Banking & Slack & Travel & Workspace \\
\midrule
Floor & .333 / .000 & .181 / .000 & .243 / .000 & .445 / .000 \\
Floor + judge & .375 / .028 & .190 / .000 & .243 / .000 & .439 / .002 \\
Floor + sanitizer & .375 / .000 & .286 / .000 & .343 / .000 & .412 / .000 \\
\textbf{SPILLWAY} (final) & \textbf{.431} / \textbf{.000} & \textbf{.286} / \textbf{.000} & \textbf{.336} / \textbf{.000} & \textbf{.382} / \textbf{.000} \\
\bottomrule
\end{tabular}
\end{table}

\paragraph{Cost}
On qwen3:8b benign traffic (Slack / Workspace), SPILLWAY calls its judge $0.43$ and $0.05$ times per episode and changes its backbone calls by less than $0.15$ (Figure~\ref{fig:position-cost} plots their sum), while MELON adds $5.29$ and $3.25$ backbone calls, Progent $8.67$ and $5.94$, and DRIFT $9.91$ and $9.59$. On the RTX~PRO~6000, SPILLWAY's episodes, timed in a separate session, were within $0.5$\,s of the undefended agent's, against $+4.9$ and $+7.9$\,s for MELON, $+1.6$ and $+2.3$\,s for Progent, and $+9.9$ and $+10.5$\,s for DRIFT.

\begin{figure*}[t]
\centering
\subfloat[Utility under attack, per suite.]{\includegraphics[width=0.380\textwidth]{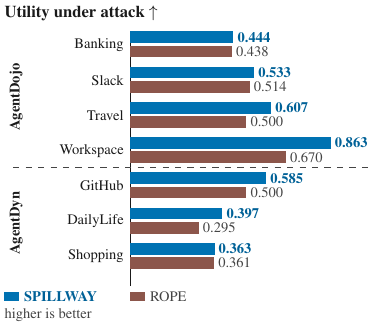}}
\hspace{0.05\textwidth}
\subfloat[In-scope attack success, per suite.]{\includegraphics[width=0.380\textwidth]{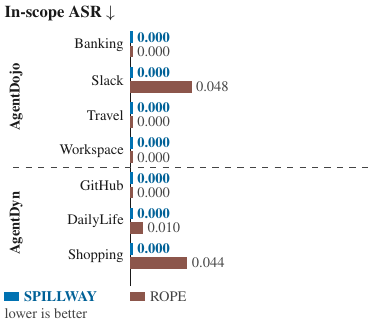}}
\caption{\textbf{Delegated-routing grid.} \textbf{AgentDyn}~\citep{agentdyn2026} and \textbf{AgentDojo} suite v1.2.2, gpt-4o-mini, standard attack, full grid. ROPE~\citep{rope2026} uses its authors' Claude Opus 4.8 router and cached scopes (Workspace scopes produced once by the same router).}
\label{tab:rope}
\end{figure*}

\paragraph{A comparison with routed-origin enforcement}
Against ROPE with its authors' Claude Opus 4.8 router (Figure~\ref{tab:rope}), SPILLWAY holds in-scope attack success at zero on all seven suites, while ROPE leaks on three (DailyLife $.010$, Shopping $.044$, Slack $.048$). SPILLWAY's utility under attack is higher on every suite, on average by $.08$ on AgentDojo ($.612$ vs.\ $.530$) and by $.06$ on AgentDyn ($.448$ vs.\ $.385$). With the agent's own model as its router (gpt-4o-mini), ROPE's mean utility falls from $.468$ to $.361$ and it still leaks on Shopping ($.072$) and Slack ($.019$). Each AgentDojo suite runs once (seed 42), and SPILLWAY's AgentDyn results average three fixed-seed runs. On benign qwen3:8b traffic SPILLWAY adds $0.19$ and $0.39$ model calls per Workspace and Slack episode, judge included, far below the model-call defenses (Figure~\ref{fig:position-cost}).

\begin{table*}[t]
\centering
\caption{\textbf{Stress tests} (suite v1.2.2): in-scope UAU~$\uparrow$/ASR~$\downarrow$. \textbf{ChatInject}~\citep{chang2026chatinject}: 665 pairs. \textbf{AutoDojo}~\citep{autodojo2026}: 389 pairs optimized against Progent, four of which SPILLWAY's rules used as development examples (held-out sets in App.~\ref{app:levels}). \textbf{PISmith}~\citep{pismith2026}: the authors' RL attacker replayed open-loop, first of ten variants per injection (ASR@1), macro over four suites. NRP~$\uparrow$/ERS~$\uparrow$ use each panel's benign utility (Table~\ref{tab:main-grid}, App.~\ref{app:levels} for ROPE on ChatInject). ROPE runs ChatInject without its clamp (App.~\ref{app:levels}). SPILLWAY in bold. $^{\ddagger}$Adapted released code (App.~\ref{app:repro}). $^{\dagger}$Authors' planner and validator.}
\label{tab:adaptive}
\label{tab:rope-chatinject}
\label{tab:pismith}
\sbox{\spwstressa}{%
\setlength{\tabcolsep}{3.6pt}
\spwtablefont
\begin{tabular}{@{}l *{8}{cc}@{}}
\toprule
 & \multicolumn{6}{c}{\textbf{ChatInject} (qwen3:8b)} & \multicolumn{8}{c}{\textbf{AutoDojo} (gpt-4o-mini)} & \multicolumn{2}{c}{\shortstack{\textbf{PISmith}\\(gpt-4o-mini)}} \\
\cmidrule(lr){2-7}\cmidrule(lr){8-15}\cmidrule(lr){16-17}
 & \multicolumn{2}{c}{Workspace} & \multicolumn{2}{c}{Slack} & \multicolumn{2}{c}{Composite} & \multicolumn{2}{c}{Banking} & \multicolumn{2}{c}{Slack} & \multicolumn{2}{c}{Travel} & \multicolumn{2}{c}{Composite} & \multicolumn{2}{c}{Macro} \\
\cmidrule(lr){2-3}\cmidrule(lr){4-5}\cmidrule(lr){6-7}\cmidrule(lr){8-9}\cmidrule(lr){10-11}\cmidrule(lr){12-13}\cmidrule(lr){14-15}\cmidrule(lr){16-17}
Method & UAU & ASR & UAU & ASR & NRP & ERS & UAU & ASR & UAU & ASR & UAU & ASR & NRP & ERS & UAU & ASR \\
\midrule
Undefended & $.052$ & $.245$ & $.095$ & $.648$ & $.269$ & $.496$ & $.444$ & $.354$ & $.524$ & $.571$ & $.329$ & $.329$ & $.342$ & $.574$ & $.128$ & $.568$ \\
\textbf{SPILLWAY} & \textbf{.421} & \textbf{.000} & \textbf{.333} & \textbf{.000} & \textbf{.463} & \textbf{.633} & \textbf{.438} & \textbf{.000} & \textbf{.562} & \textbf{.000} & \textbf{.536} & \textbf{.000} & \textbf{.621} & \textbf{.763} & \textbf{.263} & \textbf{.006} \\
MELON & $.084$ & $.020$ & $.095$ & $.000$ & $.494$ & $.660$ & $.340$ & $.021$ & $.286$ & $.000$ & $.293$ & $.000$ & $.564$ & $.716$ & $.114$ & $.061$ \\
Progent$^{\ddagger}$ & $.161$ & $.005$ & $.057$ & $.010$ & $.305$ & $.457$ & $.306$ & $.104$ & $.467$ & $.124$ & $.414$ & $.121$ & $.468$ & $.654$ & $.166$ & $.096$ \\
DRIFT$^{\dagger}$ & $.393$ & $.062$ & $.352$ & $.019$ & $.445$ & $.625$ & $.438$ & $.090$ & $.543$ & $.010$ & $.486$ & $.086$ & $.490$ & $.659$ & $.428$ & $.064$ \\
ROPE$^{\ddagger}$ & $.157$ & $.002$ & $.162$ & $.095$ & $.292$ & $.462$ & $.424$ & $.000$ & $.352$ & $.019$ & $.300$ & $.000$ & $.434$ & $.604$ & $.173$ & $.007$ \\
Spotlighting & $.191$ & $.013$ & $.076$ & $.181$ & $.398$ & $.591$ & $.431$ & $.306$ & $.543$ & $.514$ & $.343$ & $.279$ & $.360$ & $.594$ & $.151$ & $.489$ \\
PI detector & $.050$ & $.021$ & $.162$ & $.000$ & $.230$ & $.374$ & $.403$ & $.049$ & $.219$ & $.067$ & $.214$ & $.029$ & $.330$ & $.498$ & $.155$ & $.025$ \\
\bottomrule
\end{tabular}%
}
\ifdim\wd\spwstressa>\textwidth\resizebox{\textwidth}{!}{\usebox{\spwstressa}}\else\usebox{\spwstressa}\fi
\end{table*}

\paragraph{What each component adds}
Table~\ref{tab:ingress-ablation} reports component-study arms and the final system. The floor lets no attacker target through on any suite, as P1 predicts. On Slack the floor refused every attacker membership call the agent proposed. Adding the judge alone recovers Banking utility ($+.042$). Without the sanitizer, though, it admits the attacker's account on ``pay the bill'' and an attacker address as a meeting participant, which is why SPILLWAY couples the two. Adding the sanitizer instead lifts utility on Slack ($+.105$) and Travel ($+.100$), since an agent that never reads the imperative stops proposing the attacker's calls. The final system holds zero on all four suites at mean utility $.359$.

\paragraph{Open-loop replays of three recent attacks}
Beyond the standard template, we replay three recent injection attacks against SPILLWAY open-loop: a forged chat template (ChatInject, ICLR 2026, \citealp{chang2026chatinject}), optimizer-generated injections (AutoDojo, \citealp{autodojo2026}), and a released RL attacker (PISmith, \citealp{pismith2026}). ChatInject embeds a fabricated \texttt{<|im\_start|>} turn inside untrusted output. With the forged turns scrubbed, SPILLWAY has no raw hit in any of the 665 pairs and recovers part of the utility the hijack destroys (Slack $.095\!\to\!.333$, Workspace $.052\!\to\!.421$). ROPE lets the hijacked agent propose the attacker's calls and blocks most of them at execution, so its utility stays low ($.157$, $.162$) and it leaks $.002$ on Workspace and $.095$ on Slack, and DRIFT keeps more utility ($.393$ and $.352$) but leaks $.062$ and $.019$ (Table~\ref{tab:rope-chatinject}). In the attacker-as-sender variant (Appendix~\ref{app:sender}), SPILLWAY blocks the one attack that breaks the undefended agent.

AutoDojo's injections were optimized against another defended agent: Progent, the defense they were optimized against, executes them in 10 to 12\% of pairs (Table~\ref{tab:rope-chatinject}). SPILLWAY leaves no injected effect in the 389 pairs, neither a call to an attacker's target nor injected content in a final answer. Four of these pairs served as development examples for the address and payee-change rules of Sec.~\ref{sec:ingress-method}, so two held-out Banking sets (144 pairs each, Appendix~\ref{app:levels}) test the rules. Attack success is $.014$ on the set optimized against DRIFT, counting a payee transfer that AgentDojo's check misses, and $.000$ on the set optimized against the undefended agent, in which 18 of the 36 injection texts are optimized. The rules change no earlier result (Appendix~\ref{app:repro}).

\textbf{A released RL attacker.} We replay PISmith~\citep{pismith2026}, a released RL-trained attacker, open-loop against every defense: the first of its ten pair-conditioned variants per injection (ASR@1, gpt-4o-mini, Table~\ref{tab:pismith}). It breaks the undefended agent ($.568$ in-scope ASR) and five of the six baselines ($.025$ to $.489$), yet SPILLWAY holds it to a suite mean of $.006$ with more utility than every other defense except DRIFT, which leaks $.064$ (Figure~\ref{fig:tradeoff}d). ROPE comes closest at $.007$. SPILLWAY finishes nine percentage points more tasks on average ($.263$ vs.\ $.173$).

Among the defenses we run, SPILLWAY is the only one with no attacker-target success under the standard template, under ChatInject, and on AutoDojo's 389 pairs (Tables~\ref{tab:main-grid} and~\ref{tab:rope-chatinject}).

\section{Related Work}
\label{sec:related}

Information-flow and capability systems~\citep{fides2025,rtbas2025,ipiguard2025,wu2024ifc,li2026ace} constrain untrusted data. Task Shield~\citep{jia2024taskshield} checks whether an action serves the user goal.

SIEVE~\citep{liang2025sieve} checks tool transitions and argument sources against an intent graph and escalates failures. Agent-Sentry~\citep{sequeira2026agentsentry} learns a bound from benign executions and enforces it with a classifier, an allowlist, and an LLM judge. CXI~\citep{santosgrueiro2026cxi} binds field, effect, and invocation authority to an action manifest. SPILLWAY's floor needs no intent graph, learned bound, or action manifest. It reads targets from the user's task and typed read-only returns. AgentSentry~\citep{zhang2026agentsentry} counterfactually re-executes the agent, while tool-result parsing~\citep{yu2026parsing} structures the data it reads. SPILLWAY triggers no backbone re-run.

ROPE~\citep{rope2026} is closest: a router reads each request before a deterministic origin check guards sensitive parameters. SPILLWAY instead uses a flat typed-identity index and a fixed role lexicon, without that router and with lighter state than the dependency tracking of APPA~\citep{appa2026}/AgentArmor~\citep{wang2025agentarmor} and per-argument provenance of PACT~\citep{fan2026pact}. Ingress evidence bounds its selective judge. CommandSans~\citep{commandsans2025} learns a token-level sanitizer. Tool-input/output firewalls~\citep{bhagwatkar2025firewalls} ask an LLM at each boundary without checking effect targets. Their own adaptive cascade raises sanitizer attack success to $30\%$ on Slack. Agent data injection~\citep{choi2026adi} spoofs metadata inside untrusted text and defeats most tested defenses except CaMeL's strict policy. SPILLWAY's floor and sanitizer use no model. A host can use SPILLWAY's declared roles and typed fields as a fixed destination check alongside a separate semantic defense.
\section{Limitations}
\label{sec:limitations}
P1 establishes lookup soundness for attackers who write only free-text fields (\secThreat{}). In the attacker-as-sender variant, its typed sender enters $I$ and SPILLWAY blocks the one attack that breaks the undefended agent (App.~\ref{app:sender}). New tool sets require declared target roles. Delegation and semantic judgment remain targets for adaptive evaluation~\citep{zhan2025adaptive,hofer2026assessing,chen2026iterinject}. P2 confines identity-target judge recovery to eligible refusals outside removed spans.

\section{Conclusion}
\label{sec:conclusion}
SPILLWAY checks where an action may go using the user's task and typed tool returns. Targets are decided by structure, and the model is used only where structure cannot decide. The sanitizer removes matched instructions and retains evidence that bounds judge recovery. In the 12 AgentDojo cells under the standard template, only SPILLWAY recorded zero attacker-target successes, at the highest mean utility under attack and composite scores. Under ChatInject SPILLWAY had no raw hit, and under open-loop PISmith its mean in-scope ASR was $.006$. A host can audit the destination check against the task, typed fields, and declared rules. On three AgentDyn suites with declared roles, SPILLWAY recorded zero in-scope redirects at higher mean utility under attack than ROPE with its authors' router (Figure~\ref{tab:rope}), so we expect this light layer to slot into many services and agent apps.

\FloatBarrier
\section*{Acknowledgments: Generative AI Use}
Research design and preparation. The authors conceived SPILLWAY and designed its method, the threat model, the evaluation criteria, and the experimental methodology. The authors also designed and led the verification process, including repeated cross-checks of the reported results against the recorded runs. Generative AI tools, Anthropic's Claude through Claude Code and OpenAI models through Codex, supported this author-led work. They served as a discussion aid for the authors' design, generated and revised implementation code, experiment and analysis scripts, and draft manuscript text under the authors' direction, and executed the author-specified experiments through the experiment harness. The authors independently developed and validated any idea the tools raised before adopting it. AI-generated draft text and revisions appear in the Abstract, main text, and appendices. The figures were drawn by plotting and diagram code generated in the same way, every empirical plot value comes from the recorded runs, and no image-generation model was used. We disclose such outputs as AI-generated, including in workflows that combined generation with human editing or AI assistance. AI tools also helped identify candidate related work, which the authors checked before citation. The authors reviewed and revised all generated material and made all final research and editorial decisions.

Results and responsibility. The reported measurements were computed from recorded benchmark executions, including generated model outputs and released attack texts. The authors take full responsibility for the paper. Generative AI was used for editorial purposes in this manuscript, and all outputs were inspected by the authors to ensure accuracy and originality.

\label{sp:main-end}
\begingroup\raggedbottom
\bibliographystyle{IEEEtran}
\bibliography{spillway}
\endgroup

\label{sp:references-end}
\appendices
\raggedbottom
\renewcommand{\dbltopfraction}{0.95}\renewcommand{\textfraction}{0.05}\setcounter{dbltopnumber}{2}
\label{sp:appendix-start}
\section{Attack success at every scoring level}
\label{app:levels}
Table~\ref{tab:app-grid} gives, for every breadth cell, the raw and in-scope attack success of Sec.~\ref{sec:setup}. Raw applies the executed-effect rule. Collision filtering removes no hit in any cell, so the collision-filtered level equals raw. A single value means raw and in-scope agree. SPILLWAY is $.000$ at both levels in all twelve cells. Table~\ref{tab:app-autodojo} gives the \textbf{AutoDojo} panel of Table~\ref{tab:rope-chatinject} at the trace, executed, and in-scope levels. Before the address and payee-change rules, SPILLWAY's in-scope ASR on this panel was $.007$ (Travel), $.014$ (Banking), and $.010$ (Slack). After them, on two held-out Banking sets from the same release (144 pairs each, not used when the rules were written), it is $.007$ by AgentDojo's check ($.014$ counting a second transfer to an attacker-written payee) on the set optimized against DRIFT and $.000$ on the set optimized against the undefended agent, in which 18 of the 36 injection texts are optimized and the rest are the template. Utility is $.444$ ($64/144$) on each. ROPE's AutoDojo row has the setting of its gpt-4o-mini cells in Table~\ref{tab:main-grid} and uses their BU. Its ChatInject row runs without ROPE's clamp, which drops a router override looser than the audited defaults and is on in Table~\ref{tab:main-grid}. Its ChatInject composites therefore use a benign run in that setting, Workspace~$.325$ and Slack~$.286$.

\begin{table*}[t]
\centering
\setlength{\tabcolsep}{6pt}
\caption{\textbf{Appendix.} \textbf{AgentDojo} breadth grid (suite v1.2.2, 949 attacked pairs per backbone: 144 Banking, 105 Slack, 140 Travel, 560 Workspace): raw / in-scope ASR per backbone, suite, and method. One value when the two agree.}
\label{tab:app-grid}
\spwtablefont
\begin{tabular}{l cccccccc}
\toprule
Cell & Undefended & SPILLWAY & MELON & Progent & DRIFT & ROPE & Spotlighting & PI detector \\
\midrule
\multicolumn{9}{l}{\textit{Banking}}\\
\quad gpt-4o-mini & .556 & .000 & .049 & .097 & .111 & .000 & .549 & .118 \\
\quad qwen3-coder:30b & .299 & .000 & .125 & .056 & .014 & .000 & .382 & .097 \\
\quad qwen3:8b & .181 & .000 & .215 & .000 & .028 & .000 & .188 & .083 \\
\multicolumn{9}{l}{\textit{Slack}}\\
\quad gpt-4o-mini & .629 & .000 & .000 & .076 & .000 & .019 & .562 & .086 \\
\quad qwen3-coder:30b & .724 & .000 & .076 & .010 & .000 & .019 & .581 & .124 \\
\quad qwen3:8b & .448 & .000 & .457 & .019 & .010 & .019 & .467 & .086 \\
\multicolumn{9}{l}{\textit{Travel}}\\
\quad gpt-4o-mini & .300/.229 & .000 & .086/.000 & .093/.029 & .000 & .079/.000 & .236/.179 & .014 \\
\quad qwen3-coder:30b & .186/.129 & .000 & .100/.029 & .064/.007 & .000 & .043/.000 & .193/.136 & .000 \\
\quad qwen3:8b & .264/.229 & .000 & .186/.157 & .064/.036 & .000 & .043/.000 & .200/.150 & .000 \\
\multicolumn{9}{l}{\textit{Workspace}}\\
\quad gpt-4o-mini & .184 & .000 & .002 & .041 & .009 & .000 & .148 & .070 \\
\quad qwen3-coder:30b & .036 & .000 & .004 & .002 & .002 & .000 & .039 & .025 \\
\quad qwen3:8b & .020 & .000 & .011 & .004 & .014 & .000 & .020 & .014 \\
\bottomrule
\end{tabular}

\end{table*}

\begin{table*}[t]
\centering
\setlength{\tabcolsep}{6pt}
\caption{\textbf{Appendix.} \textbf{AutoDojo} on gpt-4o-mini (144 Banking, 105 Slack, 140 Travel pairs): trace-level, executed (exec.), and in-scope ASR. DRIFT's three levels coincide. ROPE's scorer reports executed and in-scope ASR, not a separate trace~($-$).}
\label{tab:app-autodojo}
\spwtablefont
\begin{tabular}{l ccc ccc ccc}
\toprule
& \multicolumn{3}{c}{Banking} & \multicolumn{3}{c}{Slack} & \multicolumn{3}{c}{Travel} \\
\cmidrule(lr){2-4}\cmidrule(lr){5-7}\cmidrule(lr){8-10}
Method & trace & exec. & in-scope & trace & exec. & in-scope & trace & exec. & in-scope \\
\midrule
Undefended & .354 & .354 & .354 & .571 & .571 & .571 & .364 & .364 & .329 \\
\textbf{SPILLWAY} & .000 & .000 & .000 & .000 & .000 & .000 & .000 & .000 & .000 \\
MELON & .021 & .021 & .021 & .000 & .000 & .000 & .043 & .043 & .000 \\
Progent & .104 & .104 & .104 & .210 & .124 & .124 & .150 & .150 & .121 \\
DRIFT & .090 & .090 & .090 & .010 & .010 & .010 & .086 & .086 & .086 \\
ROPE & $-$ & .000 & .000 & $-$ & .076 & .019 & $-$ & .014 & .000 \\
Spotlighting & .306 & .306 & .306 & .514 & .514 & .514 & .307 & .307 & .279 \\
PI detector & .049 & .049 & .049 & .067 & .067 & .067 & .029 & .029 & .029 \\
\bottomrule
\end{tabular}
\end{table*}

Table~\ref{tab:app-ropemini} gives the delegated-routing grid of Figure~\ref{tab:rope} per suite, with ROPE under its authors' router and under the agent's own model. Per-cell benign utility (BU) is in Table~\ref{tab:main-grid}.

\begin{table*}[t]
\centering
\setlength{\tabcolsep}{6pt}
\caption{\textbf{Appendix.} Delegated-routing grid (\textbf{AgentDojo} v1.2.2 above the rule, \textbf{AgentDyn} below), gpt-4o-mini: in-scope under-attack utility (UAU~$\uparrow$) / attack success (ASR~$\downarrow$), one run per suite except SPILLWAY's AgentDyn rows, which average three fixed-seed runs (144, 105, 140, 560, 180, 200, 180 pairs from top). The mean is unweighted over suites. The last two rows give the composites NRP~$\uparrow$/ERS~$\uparrow$ (Sec.~\ref{sec:setup}), macro-averaged over the seven suites. The AgentDyn suites have no benign run, so NRP and ERS here use UAU for BU, as MSB~\citep{msb2026} does for NRP (NRP $=$ UAU$\,\times(1-$ASR$)$).}
\label{tab:app-ropemini}
\spwtablefont
\begin{tabular}{l ccc}
\toprule
Suite & SPILLWAY & ROPE (Opus 4.8 router) & ROPE (gpt-4o-mini router) \\
\midrule
Banking & .444 / .000 & .438 / .000 & .431 / .000 \\
Slack & .533 / .000 & .514 / .048 & .343 / .019 \\
Travel & .607 / .000 & .500 / .000 & .350 / .000 \\
Workspace & .863 / .000 & .670 / .000 & .539 / .000 \\
\midrule
GitHub & .585 / .000 & .500 / .000 & .383 / .000 \\
DailyLife & .397 / .000 & .295 / .010 & .150 / .000 \\
Shopping & .363 / .000 & .361 / .044 & .333 / .072 \\
\midrule
Mean & .542 / .000 & .468 / .015 & .361 / .013 \\
\midrule
NRP (UAU) & \textbf{.542} & .462 & .357 \\
ERS (UAU) & \textbf{.690} & .627 & .519 \\
\bottomrule
\end{tabular}
\end{table*}

\section{Reproducibility}
\label{app:repro}
\paragraph{Agent} Every method runs on the same agent request: each benchmark's own default system prompt in the \texttt{system} role, temperature 0, seed 42, at most 8192 output tokens, and the backbone's reasoning switched off where it has a switch. AgentDojo's prompt has 469 bytes, and AgentDyn's (543 bytes) adds one line that tells the agent to complete all tasks without requesting user confirmation. Spotlighting appends its delimiter instruction. DRIFT keeps its authors' prompt, which embeds the default's instructions in an environment block and adds a line that tells its agent to complete all tasks without requesting confirmation. gpt-4o-mini-2024-07-18 is served by the OpenAI API, while qwen3-coder:30b and qwen3:8b run on Ollama. Every stack uses suite v1.2.2, whose task and data files equal AgentDojo 0.1.35's. The harness around them is each method's own, and the variant of Appendix~\ref{app:sender} re-addresses five emails. Each cell runs once, except SPILLWAY's AgentDyn cells, which average three fixed-seed runs. To bound run-to-run variation, we reran an earlier build of SPILLWAY on all three backbones and four suites across three agent-decoding seeds (\{42, 777, 123\}) with three repeats each, nine runs per backbone and 27 in all. In-scope attack success was $.000$ in every run, and per-suite utility had a standard deviation of at most $.041$ across the three repeats of a fixed seed and $.021$ across the three per-seed means.
\paragraph{SPILLWAY} The floor and the sanitizer are rules with no model call. The judge is gemma4:31b (Q4\_K\_M) on a local Ollama server with a 4096-token context at temperature 0. The address and payee-change rules change no sanitizer output on any of the tool messages behind Table~\ref{tab:ingress-ablation} and Figure~\ref{fig:position-cost}, and match no text in AgentDojo's environment data. Rerunning the recorded component-study arms with the address rules reproduces each suite's utility and attack success to the third decimal. The role lexicon has 11, 22, 10, and 11 verb prefixes for read-only, writing, destructive, and granting tools, and 19 target-argument names for AgentDojo with 6 more for AgentDyn.
\paragraph{Comparators} MELON, Progent, DRIFT and ROPE run from their released code. MELON embeds with nomic-embed-text, the prompt-injection detector is protectai/deberta-v3-base-prompt-injection-v2, and Progent's policy is written by the cell's own backbone, with Progent's own seed and retry ladder for its policy calls. ROPE's router is the agent in Tables~\ref{tab:main-grid}, \ref{tab:triple} and~\ref{tab:app-grid} and on the ChatInject, AutoDojo and PISmith panels, and its authors' Claude Opus 4.8 with cached scopes in Figure~\ref{tab:rope} and Figure~\ref{fig:tradeoff}a (with gpt-4o-mini as router in Table~\ref{tab:app-ropemini}). ROPE's clamp, off by default in its code, is on when the agent is its router, except on ChatInject, and off with its authors' Opus router.
\paragraph{Attacks and scoring} The standard attack is AgentDojo's \texttt{important\_instructions}. AutoDojo replays variant 0 of its released gpt-4o-mini injections (commit 89413642). PISmith replays variant 0 of the ten its released attacker wrote for each pair, the same texts for every defense. The AgentDyn suites have 180 (GitHub), 200 (DailyLife), and 180 (Shopping) attacked pairs. SPILLWAY's results under AgentDojo's standard attack, ChatInject, AutoDojo, PISmith, and AgentDyn, and its full row in Table~\ref{tab:ingress-ablation}, come from the final build. The other ablation arms, its cost measurements, variance runs, held-out AutoDojo sets, and sender variant come from an earlier build. Those runs and the comparators' local runs use one NVIDIA RTX PRO 6000 Blackwell GPU.

\section{Attacker-as-sender variant}
\label{app:sender}
The variant tests SPILLWAY outside P1's premise: the five injection-carrying Workspace emails come from the attacker's own address, so the address enters $I$ through the typed sender field. It covers the nine user tasks that read these emails and all 14 injection tasks (126 pairs, qwen3:8b). SPILLWAY, the floor-only arm, Progent, and the prompt-injection detector let no attack through. The undefended agent, MELON, Spotlighting, and DRIFT each execute the attack in one of the 126 pairs. In that pair the floor-only agent tried to email the attacker's address and was refused, because the injected text repeats the address and the floor refuses a target repeated in free text. With the sanitizer on, the agent never proposed the call. Utility under attack is $.548$ for SPILLWAY, $.611$ for the floor-only arm, and $.603$ undefended.

\end{document}